\documentclass{aa}

\usepackage{graphicx}
\usepackage{pstricks}

\usepackage{txfonts}
\usepackage[authoryear,square]{natbib}
\usepackage[switch,pagewise,running]{lineno}

\usepackage{color}

  \usepackage{natbib}
  \bibpunct{[}{]}{;}{a}{}{,}

\usepackage[colorlinks=true]{hyperref}
\hypersetup{
    bookmarks=true,         % show bookmarks bar?
    unicode=false,          % non-Latin characters in Acrobat's bookmarks
    pdftoolbar=true,        % show Acrobat's toolbar?
    pdfmenubar=true,        % show Acrobat's menu?
    pdffitwindow=false,     % page fit to window when opened
    pdftitle={The trans-neptunian Haumea},
    pdfauthor={C. Dumas, B. Carry, D. Hestroffer and F. Merlin}, 
    pdfsubject={Planetary Science}, 
    pdfkeywords={},         % list of keywords
    pdfnewwindow=true,      % links in new window
    colorlinks=true,        % false: boxed links; true: colored links
    linkcolor=gray,         % color of internal links
    citecolor=blue,         % color of links to bibliography
    filecolor=gray,         % color of file links
    urlcolor=gray           % color of external links
}

\begin{document}

\title{High-contrast observations of 136108 Haumea%
  \thanks{Based on observations
    collected at the European Southern Observatory, Paranal, Chile -
    \href{http://archive.eso.org/wdb/wdb/eso/sched_rep_arc/query?progid=60.A-9235(A)}{60.A-9235}}}
\subtitle{A crystalline water-ice multiple system}

\author{C. Dumas
  \inst{1}
  \and
  B. Carry\inst{2,3}
  \and
  D. Hestroffer\inst{4}
  \and
  F. Merlin\inst{2,5}
}

\institute{European Southern Observatory.
  Alonso de C\'ordova 3107, Vitacura, Casilla 19001, Santiago de Chile, Chile\\
  \email{cdumas@eso.org}
  \and
  LESIA, Observatoire de Paris, CNRS. 
  5 place Jules Janssen, 92195 Meudon CEDEX, France\\
  \email{benoit.carry@sciops.esa.int}
  \and
   European Space Astronomy Centre, ESA. 
   P.O. Box 78, 28691 Villanueva de la Ca\~{n}ada, Madrid, Spain
  \and
  IMCCE, Observatoire de Paris, CNRS.
  77, Av. Denfert-Rochereau, 75014 Paris, France\\
  \email{hestro@imcce.fr}
  \and
  UniversitŽ Paris 7 Denis Diderot.
  4 rue Elsa Morante, 75013 Paris, France\\
  \email{frederic.merlin@obspm.fr}
   }
   \date{Received 2010 May 18 / Accepted 2011 January 6}

\abstract
{
  The trans-neptunian region of the Solar System is populated by a
  large variety of icy bodies showing great diversity in orbital behavior,
  size, surface color and composition. One can also note the
  presence of dynamical families and binary systems.
  One surprising feature detected in the spectra of some of
  the largest Trans-Neptunians is the presence of crystalline water-ice. This is
  the case for the large TNO (136\,108) Haumea
  (2003\,EL$_{61}$).}
{We seek to constrain the state of the water ice of Haumea and its
  satellites, and investigate possible energy sources to maintain the water ice in its crystalline form.} 
{Spectro-imaging observations in the near infrared have been performed with the
  integral field spectrograph SINFONI mounted on UT4 at the ESO Very
  Large Telescope.
  The spectra of both Haumea and its larger satellite Hi'iaka are
  analyzed.
  Relative astrometry of the components is also performed, providing a check of the orbital solutions and equinox seasons.}
{We describe the physical characteristics of the crystalline water-ice present on the surface of Haumea and its largest satellite Hi'iaka and analyze possible sources of heating to maintain water in crystalline state: tidal dissipation in the system
  components \textsl{vs} radiogenic source. The surface of Hi'iaka
  appears to be covered by large grains of water ice,
  almost entirely
  in its crystalline form. Under some restricted conditions, both radiogenic heating and tidal forces between Haumea and Hi'iaka could provide the energy sufficient to maintain the ice in its crystalline state.}{}

  \keywords{Kuiper belt objects: individual: Haumea --
    Methods: observational --
    Techniques: high angular resolution --
    Techniques: imaging spectroscopy --
    Infrared: planetary systems}
  \maketitle

%
%________________________________________________________________

%\pagewiselinenumbers

\section{Introduction}
  The planetesimals orbiting beyond Neptune, the transneptunian objects
  (TNOs), are the remnant of the Solar System formation in its outer
  part. They are thought to be among the most pristine objects of our solar system, although
  their outer surface layers have been altered by irradiation and collisions over the age of the solar system. 
 Currently the TNOs population accounts for $\sim$1300 known objects, which are difficult to observe due to 
 their extreme heliocentric distances and relatively small size. 
  Our knowledge of their physical characteristics is for now limited to studying the few largest and brightest objects, which still   reveal that this population displays a large fraction of binary and multiple systems when compared  to other small solar 
  system bodies such as main belt asteroids \citep[e.g.][]{noll08}. 
  Trans-neptunian binaries can be found as gravitationally bound systems with similar mass components, but systems harboring 
  smaller moons, which are by definition harder to detect, have also
  been discovered around (134\,340) Pluto, (136\,108) Haumea  (2003\,EL$_{61}$),  (50\,000) Quaoar, (90\,482) Orcus, 
  \textsl{etc} \citep{noll08}.
  Thanks to their binary nature, the total mass of these systems
   can be inferred, which, when combined to spectroscopy and radiometric sizes, 
  provides a valuable tool to characterize their surface and internal physical properties.
  
  Haumea is  the largest member of a TNO family, likely the outcome of a collision \citep{brown07, ragozzine07, schaller08, rabinowitz08, snodgrass10}. Here we report spectro-imaging observations of all three components of 
  the Haumea system performed in 2007 at the ESO Very Large Telescope. Our data and related compositional modeling show that 
  the surface of the outer satellite Hi'iaka is mostly coated with crystalline water-ice, as in the case of the central body Haumea 
  \citep{trujillo07, merlin07, pinilla09}.  We also discuss the effects of tidal torques 
  as a possible source of energy responsible for the crystalline state of the water-ice of Hi'iaka.

\section{Observations and data-reduction}
  Haumea was observed in H and K bands on March 15 UT 2007, using the laser guide star facility (LGSF) and the SINFONI 
  instrument (Spectrograph for INtegral Field Observations in the Near Infrared), both installed at the 8m ``Yepun'' unit of 
  the ESO Very Large Telescope. The use of SINFONI for the observations of the large TNOs Haumea and Eris has been described 
  in earlier papers \citep{merlin07, dumas07} and more information about this instrument can be found
 in \citet{eisenhauer03} and \citet{bonnet04}. In a nutshell, SINFONI is an integral field spectrometer
  working in the [1.0-2.5] $\mu$m range, also equipped with an adaptive optics (AO) system with Natural Guide Star (NGS) 
  and Laser Guide Star (LGS) channels. While our previous published observations were obtained in non-AO mode (seeing limited), 
  the results presented in this paper make use of the AO system and the LGS facility. The laser produces an artificial visible-light 
  star of R$_{mag}\sim$13.4 in the line of sight of Haumea (V$_{mag} \sim $17.4), returning thus a gain of 4
  magnitudes for characterizing the higher orders of the wavefront in comparison to non-laser observations. Haumea itself was 
  used as a reference source for the tip-tilt, delivering optimal correction by the AO-LGS system. The atmospheric conditions 
  were extremely good during the observations, with an uncorrected  seeing varying between 0.5 and 0.6\arcsec. 
  On 3/15/2007, between 6h34 UT and 7h24 UT, six exposures of 300s each were obtained on Haumea (total integration time of 
  1h), inter-spaced by 3 exposures of 300s to record the sky background. We used the H+K spectral grating  (spectral resolution 
  of $\sim$ 1500) covering both H and K bands simultaneously, and a plate scale of 100 mas/spaxel 
  (3\arcsec $\times $ 3\arcsec total field). Calibrations to correct our spectra from the solar response and telluric absorption 
  features were obtained immediately after Haumea by observing the local telluric standard HD~142093~(V$_{mag}\sim7.3$, G2V) 
  in NGS mode at similar airmass and with the same instrumental setting. \\ 
  \indent The data (science target and telluric standard) were mainly
  reduced  using the ESO pipeline 1.9.3  \citep{mod06}
  We first corrected
  all raw frames from the noise pattern of dark and bright horizontal
  lines introduced when reading the detector.
   We then used the ESO pipeline to produce all master files needed by the data reduction 
  such as the badpixel masks, master darks
  and flats, as well as the wavelength and distortion calibration files,
  which respectively associate a wavelength value to each pixels,
  and reconstruct the final image cubes. Each object frame was
  subtracted from the sky frame recorded closest in time and the quality
  of the sky subtraction was improved by enabling the correction of sky
  residuals in the pipeline, \textsl{i.e.} by subtracting the median value of each
  image slice in the reconstructed, sky-corrected, spectro-image cube. \\ %
\begin{figure}
  \includegraphics[width=9cm]{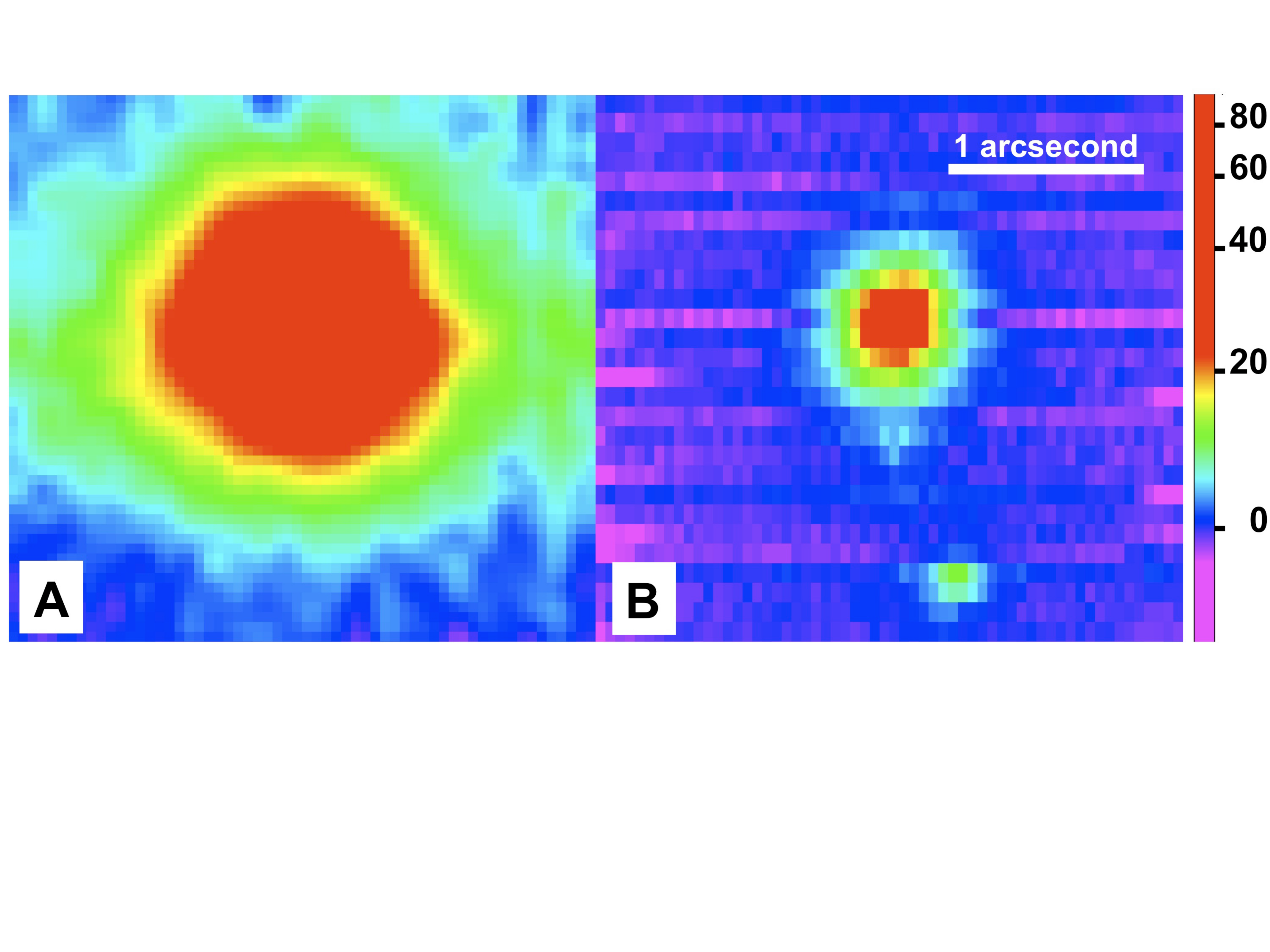}
  \vspace*{-2.5cm}
  \caption{Comparison of H+K band SINFONI images of Haumea obtained under similar conditions but in seeing limited observations (A, left) and LGS-AO corrected mode (B, right). The spatial and intensity scales are similar and the intensity is given in ADU. The improved contrast and spatial resolution of the AO image (B) is readily apparent in comparison to the no-AO image (A), making possible the detection of the two faint satellites: Namaka (faintest, just below Haumea) and Hi'iaka (brightest, bottom of image).  The images were obtained by summing all the slices of our data-cube to produce the equivalent of a broad H+K band image.  
  }
  \label{fig: cube}
\end{figure}
  Fig.~\ref{fig: cube} shows two H+K -band images of Haumea obtained in seeing
  limited and LGS modes. The improvement
  in contrast returned by the LGS is immediately apparent, as the two
  satellites of Haumea are visible in the LGS image, allowing us to
  carry out a detailed spectroscopic and astrometric study of the
  components of this system. \\ 
  
  We thus were able to extract separately the spectra of Haumea and its brighter
  satellite Hi'iaka. The faintest satellite Namaka could not be spectrophotometrically isolated from Haumea
   due to its too close proximity 
  at the time of these observations. Nevertheless we could neglect the contribution of the satellite to the overall 
  spectrum as its H~magnitude is $\sim$24.9 \citep{fraser09}, \textsl{i.e.} within the detection level for a given wavelength bin of our data cube.  The individual spectra were then corrected from the
  remaining bad pixels, combined and finally divided by the spectrum of
  the local telluric standard HD~142093. A detailed analysis of the
  cube and subsequent modeling of the spectra
revealed that division of our spectra by the solar analog  had the
effect of introducing  a small artifact in the spectrum of Haumea in
the [1.65-1.8] $\mu$m range. This particular feature was due to the contamination of our spectra by 
a faint background object within the close vicinity of the standard star.
We characterized the impact of the contaminant by dividing our
spectrum of HD~142093 with the spectrum of good solar analog,
HD~11532~(V$_{mag}\sim9.7$, G5) used by our ESO Large Program
(Prog.ID 179.C-0171, PI: Barucci) and obtained with a similar setup and airmass
($\Delta_{airmass}\sim0.03$). We then applied correction to our final
spectra of Haumea and Hi'iaka by dividing both of them by the relative
response of the two telluric standard stars over the [1.65-1.8] $\mu$m range.

\section{Structure of the water ice}
\subsection{Spectral behaviour}

\begin{table}
  \centering
  \begin{tabular}{ccc}
     \noalign{\smallskip}
        \noalign{\smallskip}
    \hline\hline
Absorption band &	Band depth (Primary) &	Band depth (Satellite)\\
  \hline
1.50 $\mu$m	& 0.36 $\pm$ 0.05	& 0.53 $\pm$0.25\\
1.65 $\mu$m	& 0.24 $\pm$0.04	& 0.54 $\pm$0.25\\
2.00 $\mu$m	& 0.55 $\pm$ 0.05	& 0.72 $\pm$ 0.35\\
    \hline
  \end{tabular}
  \caption{Depth of the water ice absorption bands in the spectra of Haumea and its outermost satellite Hi'iaka.} 
  \label{tab: banddepth}
\end{table}

\begin{figure}
  \hskip-1cm\includegraphics[width=11.5cm]{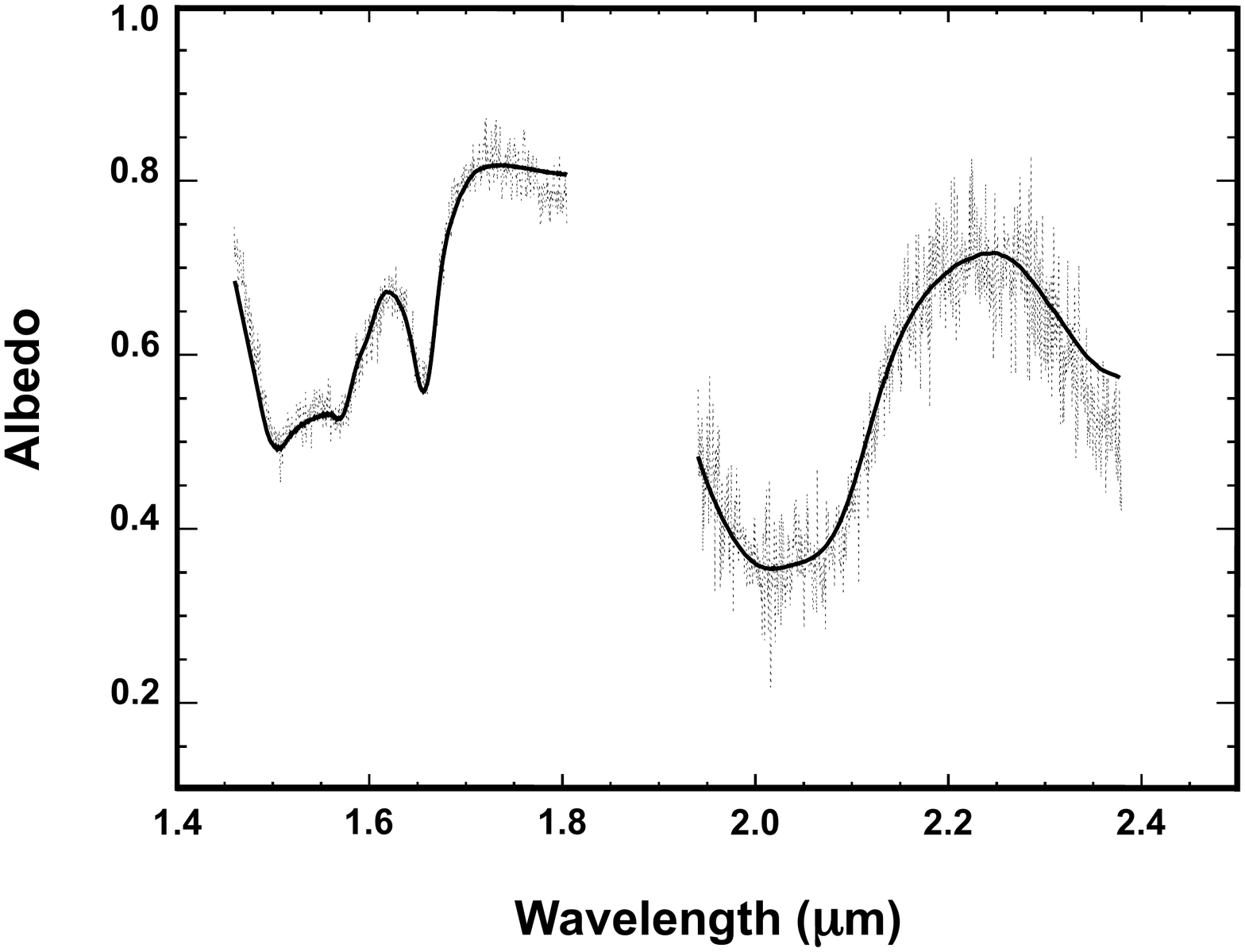}
  \caption{Spectrum of Haumea (thin dashed line) obtained with the SINFONI
    instrument using LGS-AO assisted observations. The 1.65 $\mu$m
    feature of the crystalline state of water ice is clearly seen in
    our data. From Hapke modeling (thick line) we derive  that the
    surface is made of a mixture of 73\% (particle size of 9 $\mu$m)
    of crystalline water ice, 25\% (particle size of 10 $\mu$m) of
    amorphous water ice and 2\% (particle size of 10 $\mu$m) of a dark
    compound such as Titan Tholin. No other major compound seem to be
    present on the surface of Haumea. 
  }
  \label{fig: haumeaspec}
\end{figure}

\begin{figure}
  \hskip-1cm\includegraphics[width=11.5cm]{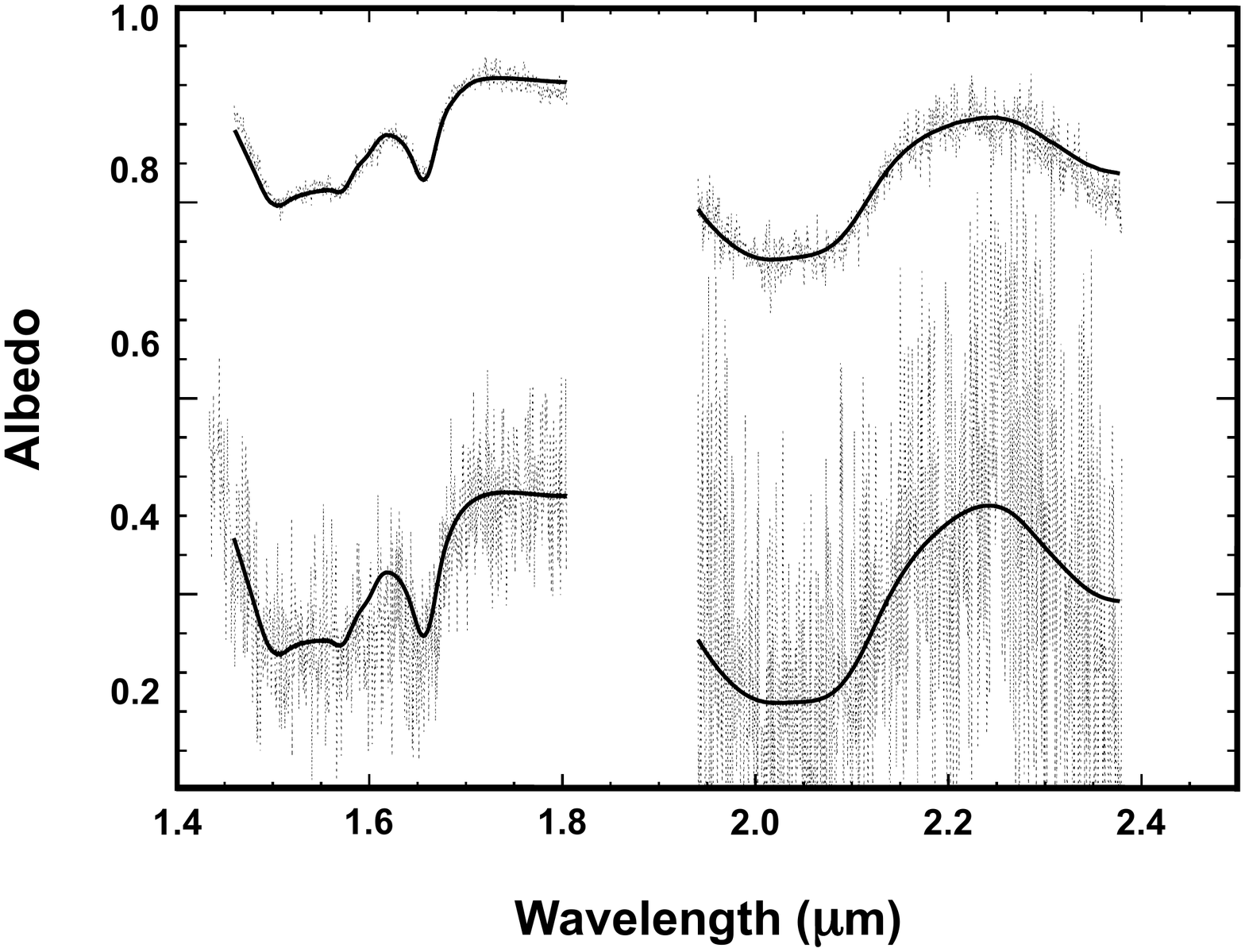}
  \caption{Spectra of Haumea (top, offsetted by +1 unit for clarity) and of Hi'iaka (bottom), 
  the largest of Haumea's satellites. The spectrum of the satellite 
    (thin dashed line) was extracted from the same data set than Fig.~\ref{fig:
      haumeaspec}. The 1.65 $\mu$m feature of crystalline state of
    water ice is also clearly seen and the band appears to be deeper
    for the satellite than for the central body. Hapke modeling (thick
    line) suggests that the surface of the satellite is nearly
    uniquely made of  crystalline water ice with larger particle size
    (20 $\mu$m) than Haumea. These results support a surface less
    altered than that of Haumea.  
  }
  \label{fig: haumeaspecsat}
\end{figure}

Our spectra of Haumea (Fig.~\ref{fig: haumeaspec}) and of its
brightest satellite (Fig.~\ref{fig: haumeaspecsat})  reveal clear
absorption bands of water ice as reported by \citet{barkume06} around
1.5 and 2.0 $\mu$m. Previous reports \citep[e.g.][]{trujillo07,
  merlin07} had also shown that the spectrum of Haumea displays the
clear signature of crystalline water ice at 1.65 $\mu$m. Here, these
LGS-assisted VLT observations clearly show that water ice in its
crystalline state is similarly present on the brightest of the
satellite. Crystalline ice on Hi'iaka was also reported  previously by
Takato et al. (unpublished) from seeing-limited observations carried
out at Subaru under very good atmospheric conditions, while this paper
reports LGS-assisted observations of Haumea's satellite and Hapke
modeling of its reflectance spectrum. The primary object spectrum does
not display other major absorption bands in the 1.45-2.35 $\mu$m range. We can suspect a couple of absorption bands around 2.21 $\mu$m and 2.25 $\mu$m in the spectrum of Haumea, which, if real, could maybe be explained by the presence of NH$_3\cdot$H$_2$O and, tentatively, NH4$^{+}$, on surface, the latter being the likely product of irradiation of ammonia hydrate \citep{cook09}; though better data are needed to confirm these bands. The spectrum of Hi'iaka is still too noisy to search for the signature of any additional compounds. 

The crystalline water ice band (at 1.65 $\mu$m) is very deep in the
spectrum of both objects. This behaviour is similar to those of pure
crystalline water ice at low temperature
(\citep[see][]{grundy98}). Considering the high albedo of the primary
object \citep{stansberry08}, we can assume that crystalline water ice
is the major compound on the surface of Haumea and Hi'iaka (and likely
Namaka as well). For both spectra, we analyzed the relative depth of the water ice absorption
bands at 1.5 $\mu$m, 1.65$\mu$m and 2.0x$\mu$m in comparison to the continuum flux estimated at each band center.
 This continuum was first divided out from our measured spectrum before estimating the depth of each band. 
 The results are given in Table~\ref{tab: banddepth}. 

 The absorption bands of the spectrum of the satellite are deeper than
 those of the spectrum of Haumea by a factor $\sim$1.5 for the wide
 absorption bands (1.5 and 2 $\mu$m) and more than 2.5 for the finest
 1.65 $\mu$m band. This implies larger grain size of the water ice on
 the surface of the satellite. Concerning the depth of the absorption
 band at 1.65 $\mu$m, we can suggest that the surface of the satellite
 has less suffered from the irradiation processes than the surface of
 Haumea (see \citet{merlin07} for a discussion about the shape and
 location of the 1.65 $\mu$m band).

\subsection{Spectral modeling}
To investigate the surface properties of Haumea and Hi`Ôiaka, we ran a
radiative transfer model, based on Hapke theory \citep{hapke81}. We computed the
geometric albedo at zero phase angle from Eq. (44) of \citet{hapke81}. The
phase function, that describes the angular distribution of light scattered
from a body, is represented by a single Henyey-Greenstein function \citep{henyey41}
with an asymmetry parameter of v=-0.4. The
backscattering parameter is B= 0.7. These values are close to those used
by \citet{verbiscer98} for the icy satellites of the giant
planets, which exhibit similar strong water ice features. We follow the
formalism of \citet{emery04} to compute the geometric albedo from
different compounds, assuming a Ósalt and pepperÓ or an ÓintimateÓ
mixture. The free parameters of our models are the grain size and the
relative amount of each chemical compound. The lowest reduced $\chi$$^{2}$ values
between the observed spectra and our synthetic spectra were reached using
the Marqvardt-Levenberg algorithm, although it is important to note that
our model results are not unique and only show the most probable surface
composition from our initial set of probable chemical analogs \citep[see][on a discussion of the limits of this model]{barucci08a}.

To perform our spectral modeling, we used optical constants of several
ices at low temperature (close to 40 K) suspected to be present on the
surface of these icy bodies such as pure and amorphous water ice \citep{grundy98},
as well as pure methane ice \citep{quirico97}. We also
used optical constants of dark compounds such as amorphous black carbon
\citep{zubko96} and Titan Tholin \citep{khare86}, which reproduce
the low albedo of a large portion of these objects \citep{stansberry08}. 

Our best result, obtained assuming an albedo of 0.6 (normalized over
the 1.6-1.7 $\mu$m region of the spectrum), includes 73\% (particle
size of 9 $\mu$m) of crystalline water ice, 25\% (particle size of 10
$\mu$m) of amorphous water ice and 2\% (particle size of 10 $\mu$m) of
Titan Tholin. The albedo value in the near infrared was determined
from its V albedo \citep{stansberry08}, its V-J colour and the
reflectance ratio reported between the CH$_4$ band at 1.6 $\mu$m
(CH4s) and J band \citep{lacerda08}. We have normalized our spectra by
convolving them with the response curve of the CH4s filter used by
\citet{lacerda08} in the H-band region. For the satellite, we treated
its albedo as a free parameter in our model and the best fit was
obtained for an albedo of 0.4 in CH4s band and a composition made of
100\% of crystalline water ice  (particle size of 20 $\mu$m). The
  results of our spectral modeling are given in Figure~\ref{fig:
    haumeaspec} and Figure~\ref{fig: haumeaspecsat}.

\subsection{Discussion}

Our observations and modeling results clearly show that crystalline
water ice is present on the surface of the largest satellite, and
likely more abundantly than on the surface of the central body (larger
particle size and greater amount). Even though our results would
require an independent determination of Hi'iaka's albedo, it is highly
probable that the surface of the satellite is completely covered by
crystalline water ice, especially if the exact albedo in the CH4s band
is close to, or even larger than, the value of 40\%. The presence of
crystalline water ice on the surface of Hi'iaka demonstrates that
crystalline ice can be present on the surface of very small
bodies. Indeed, if we adopt a size of a 1600 km diameter for the
primary \citep{rabinowitz06}, a similar visible albedo between the two
bodies, and a magnitude difference of 3.3 \citep{brown05}, we derive a
diameter of 170 km for the largest satellite. 

\citet{mastrapa06} and \citet{zheng08} have shown that the crystalline
water ice feature almost disappear after irradiation over a time span
of only several Myr to several hundreds Myr, hence the life time of
the crystalline state of water ice is expected to be small on outer
solar system objects, especially for low temperature surfaces. Also,
crystalline water ice can only be formed from amorphous water ice
after episodes of sufficient heating, this mechanism being very
efficient above 100 K \citep{jewitt04}, but still possible at lower
temperatures. Based on this, some competitive mechanisms must be
involved to explain that water ice is found mostly in its crystalline
state over planet satellites and TNOs, including those of small size.
As shown here our spectral modeling results show that crystalline
water ice is dominant and ``fresh'' (less than several Myr) on the
surface of Haumea and its largest satellite.  
\citet{zheng08} showed that the amorphization of crystalline ice by
irradiation becomes less efficient with increasing temperature, the
effect of ``thermal recrystallization'' becoming even dominant at
higher temperatures than 40K. This could partly explain why
crystalline ice is still found on small outer solar system bodies.
Several authors proposed also cryovolcanic processes to explain the
observation of crystalline water ice \citep[e.g.][]{jewitt04,
  cook07}. From observations, this assumption could be possible for a
few objects where ammonia ice has been detected; for instance: Charon
\citep{brown00, buie00, dumas01, cook07}, Quaoar \citep{jewitt04}, or
Orcus \citep{barucci08b}, as ammonia depresses the melting point and
could cause liquid to be compressed and pressurized enough at high
depth to reach the surface \citep{cook07}. However, the presence of
absorption bands due to ammonia is not definitive in our spectra.  

\citet{brown07} have reported the probable discovery of a family of
Kuiper belt objects with surface properties and orbits that are nearly
identical to those of Haumea, likely fragments of the ejected ice
mantle the parent body. Recent simulations performed by
\citet{ragozzine07} seem to confirm this hypothesis even if the epoch
found for the collision seems too ancient (1Gyr) to conserve the fresh
mantle of these bright objects. From photometry, \citet{rabinowitz08}
show that the members of this family have common phase curve and have
the bluest colour among all the TNOs. These observations suggest a
high albedo for all of the objects and assume very fresh
surfaces. \citet{barkume08} show that all observed members of this
family show clear absorption features of crystalline water ice, not
observed in other small TNOs (diameter smaller than $\sim$1000 km),
although the number of putative family members has recently been
lowered by \citet{snodgrass10}. Still, the hypothesis of an energetic
collisional event could provide a scenario explaining the presence of
nearly pure crystalline water ice on the surface of Hi'iaka,
especially if we consider that the largest of Haumea satellite is
still too small to reach melting point of H$_{2}$O at any depth. The
next section below investigates the possibility of maintaining
interior temperatures high enough by involving other scenarios:
radiogenic heating and tidal effects between Haumea and Hi'iaka.

\section{Orbits of Haumea's satellites}

\begin{figure}
\centerline{
  \includegraphics[width=\hsize]{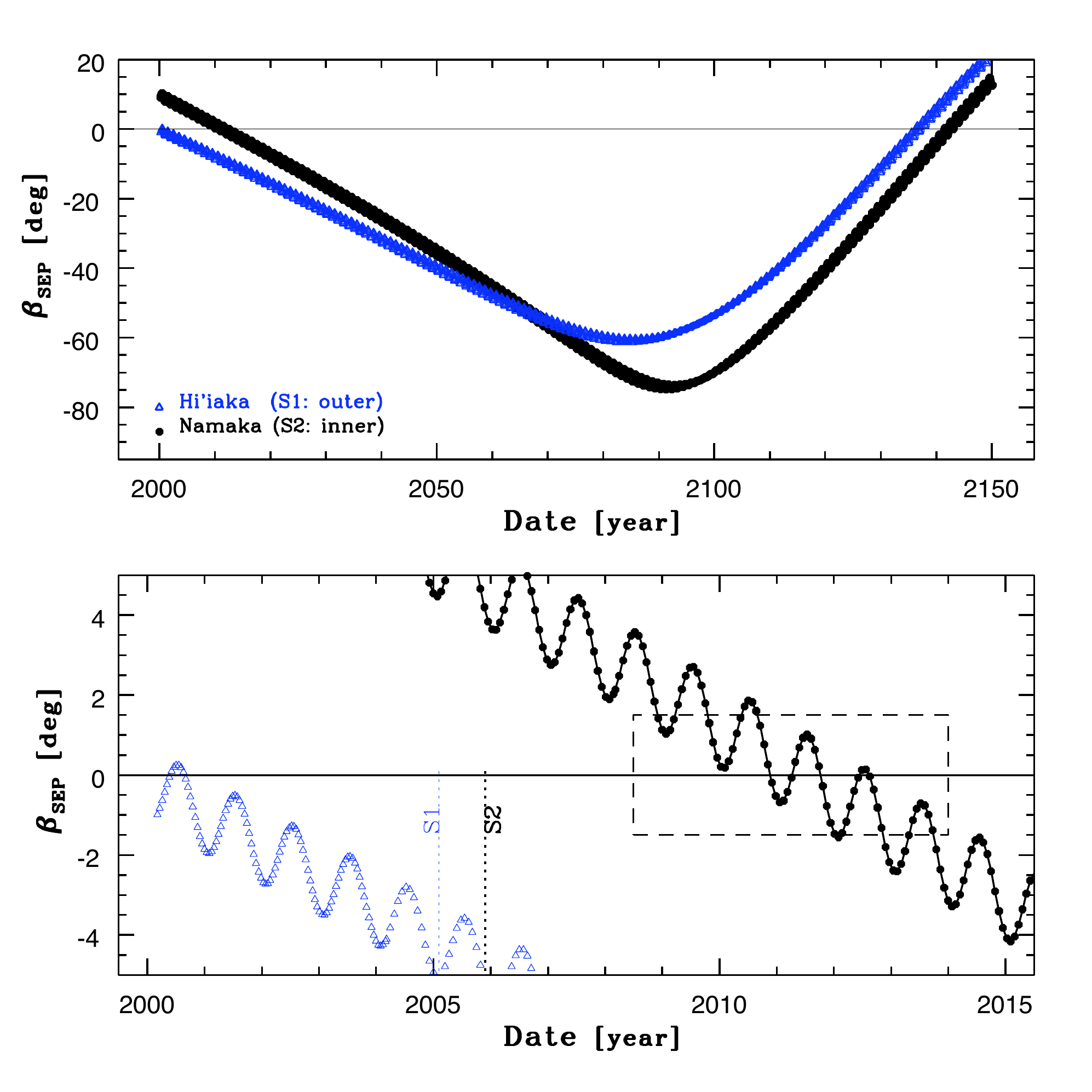}
}
\caption{Prediction of the Earth elevation (sub Earth point's latitude:
  SEP$_\beta$) above 
  the orbital plane for each of the two satellites. In open triangles
  (blue) for the outer satellite S1 Hi'iaka discovered in 2005; in filled
  circles (black) for the lately discovered inner satellite S2 Namaka. There
  are two equinox seasons during one orbital period of the system around
  the Sun, hence approximately one each 130 years. Equinox season for
  the inner satellite have started and last for about 2 years. The 
  characteristic modulation of the curve is the effect of the parallax, 
  yielding several occultations periods. Note that the curve only 
  depends on the inclination of the orbit.}
\label{fig: equinox}
\end{figure}

  \indent Taking advantage of the imaging capabilities of SINFONI, we
  extracted the relative astrometric positions of (136\,108) Haumea and its
  two satellites by fitting Gaussian profiles on each of the components.
  We found the brighter satellite (S1: Hi'iaka) at
  ($-$0.\arcsec277$^{\pm.01}$ E,
   $-$1.\arcsec318$^{\pm.01}$ N) 
  from (136\,108) Haumea,
  and the faintest satellite (S2: Namaka) at
   ($+$0.\arcsec026$^{\pm.02}$ E,
   $-$0.\arcsec528$^{\pm.03}$ N).
  The large error bars along the South-North direction are due to the
  non-squared shape of SINFONI spaxels, which are twice as large in the SN 
  direction than along the EW direction. \\
  \indent These positions agree with orbits recently determined by \citet{ragozzine09}, 
 emphasizing the astrometric quality of the data obtained with SINFONI at the VLT.  
  These orbits lead to mutual events (eclipses) between Haumea and its
  satellite Namaka within the period $\approx$ 2008-2011
  (see Fig.~\ref{fig: equinox}).
  Such events are of prime importance as their photometric follow-up
  can lead directly to a direct determination of the real
  size of the components, and hence their bulk density, with high
  accuracy. 
  These  events have been indeed predicted by \citet{fabrycky08} based on
  HST and Keck observations. Due to the parallax of the system, the Earth will cross several times 
the orbital plane of the inner satellite, leading to several 
favorable opportunities to observe transits and occultations phenomena. 
It is worth noting that some symmetric solutions for the orbit orientation, although less likely, cannot strictly be ruled out, which would have strong consequences on the prediction of the equinox seasons.
Because of scarcity of such occultation events within the transneptunian region, it is important to 
gather additional astrometric data on the position of the inner satellite for better 
constraining its orbit orientation, and consequently the prediction of the mutual events.

\section{Discussion}
This paper shows that crystalline ice is not only present on the
  largest body of the Haumea multiple system, which could be explained by 
 the long live effect of radiogenic heating, but also on the external satellite Hi'iaka (and hence likely
 on the inner satellite Namaka).  
 Several mechanisms to explain the widespread presence of crystalline ice among primitive small solar system objects have already been proposed \citep{jewitt04, grundy06, cook07}, and all require that some earlier heating events above 80-90\,K  \citep{schmitt88} might have occurred at large heliocentric distances.  In the following, we explore the efficiency of radiogenic heating and  tidal dynamical effects as possible heat sources for maintaining water ice in its crystalline state over Hi'iaka. \\

\begin{table*}
  \centering
  \begin{tabular}{cccccccc}
    \hline\hline
    & $m^{(a)}$ & $r^{(a)}$ & Si
    & E$_{\textrm{Si}}$ & E$_{\textrm{Tides}}$
    & $\delta$T \\
    & (kg) & (km) & (\%) & (W)& (W) & (K.s$^{-1}$)\\
    \hline
    Haumea  & $4.00 \times 10^{21}$ & 690 &  88--97  & $\approx 10^{10}$ &
    $\approx 5 \times 10^{9}$ & $6 \times 10^{-16}$\\
    Hi'iaka & $1.8  \times 10^{19}$ & 195 & $\leq$10 & $\approx 10^{7}$  &
    $\approx 10^{7}$          & $3 \times 10^{-16}$\\
    Namaka  & $2    \times 10^{18}$ & 100 & $\leq$10 & $\approx 10^{6}$  &
    $\approx 10^{7}$        & $6  \times 10^{-15}$ \\
    \hline
  \end{tabular}
  \caption{Orbital and physical parameters of the Haumea's system:
In addition to the mass ($m$) and size ($r$) of Haumea and its
satellites, we report the silicate content in mass (Si) estimated from
the body's density, and the subsequent radiogenic energy
(E$_\textrm{Si}$), as well as the total energy available from tidal
dissipation (E$_\textrm{Tides}$). Last column reports the resulting
temperature rate. (a) is from \citet{ragozzine09}. 
}
\end{table*}

  The radiogenic heating still present in such bodies comes from long
  period unstable elements such as  $^{40}$K, $^{232}$Th and
  $^{238}$U. The heating depends on the volume/surface ratio of the body
  and mostly the total mass of silicates (one has globally
  $4.5\times10^{-12}\,$W.kg$^{-1}$ for rocks).
  Assuming a two layers model for the internal structure of Haumea,
  with a silicate core (density of 3,500 kg.m$^{-3}$) and a water ice
  surface (density of 900 kg.m$^{-3}$), we find that the rock fraction
  should represent between 88\% and 97\% of the mass of Haumea
  \citep[4.00 $\pm$ 0.04 $\times 10^{21}$
    kg,][]{ragozzine09} to comply with the possible density
  range 
  (2,600 to 3,300 kg.m$^{-3}$) reported by
  \citet{rabinowitz06}.
  The radiogenic energy presently available is thus of the order of 
  10$^{10}$\,W, about 10 times what is expected for asteroid (1)
  Ceres. 
  In the perspective of a great impact scenario, the fraction of
  rock in the satellites should be lower. Still, a 10\% rock-fraction for the satellites
  would provide a radiogenic energy of the order of
  $10^{6}$\,W for Namaka, and $10^{7}$\,W for Hi'iaka, which is
  comparable to the icy satellite of Saturn Tethys (although it
    is understood that in the case of Tethys, resurfacing processes to
    maintain water ice in its crystalline state differ, and invoke the
    action of particle bombardment from the nearby E-ring and Saturn's
    magnetosphere).  \\ 

  \indent Assuming now that all the energy involved from the tidal flexure is
  dissipated, an {\em upper-bound} to the crystalline-ice production can
  be obtained.  
  The amount of thermal energy dissipated in the body is the residual of
  the transfer of orbital and rotational energy of the deformed body
  $\dot E_{\rm th}=|\dot W_{\rm\scriptstyle tide} + \dot
  W_{\rm\scriptstyle rot}|$. 
  Here we have to consider the more general case of inclined and
  eccentric orbits, elongated primary (the large lightcurve amplitude
  observed
  \citep{rabinowitz06, lellouch10}
  is clearly associated to shape effect, although the object displays albedo markings
  \citep{lacerda08}), and non
  synchronous rotation.
  However, for the upper-bound computation derived here, we will neglect the effect of eccentricity, obliquity and shape, 
  according to the considerations proposed by \citet{ferraz08} for fast rotators\footnote{This simplification is based on the following consideration: If the very elongated primary is a fast rotator, the tides it will generate will be of high  frequencies 
  (see  \citet{ferraz08} for more detailed information).}. 
  Hence, the energy released by the effect of tides can be expressed as:
\begin{equation}  
  \dot E_{\rm th} = \frac{3}{2} ~
                    \mathcal{G} m^2
                    \frac{r^5}{a^6} ~
                    \Omega ~ k_d . \epsilon,
\end{equation}
  \noindent where $a$ is the semi-major axis of the orbit;
  $\Omega$,
  $m$ and $r$ Haumea's spin rate, mass and equivalent radius;
  $k_d$ the Love dynamical number;
  $\epsilon$ the phase lag of the tide
  and $\mathcal{G}$ the constant of gravitation.\\
  \indent Considering the rheology of the material  (viscosity,
  elasticity and 
  rigidity) and response to forced periodic oscillations through
  $k_d . \epsilon$,
  only a fraction of this energy will be dissipated.
  For a typical icy body one can
  assume $Q\approx$ 30-100,
  which also depends on the temperature of the
  ice. The Love number scales linearly with the body's size and one can
  assume $k_d . \epsilon \propto 10^{-2}$ for Haumea, and
  $\propto 10^{-3}$ 
  for its satellites.
  Thus, the total energy 
  dissipated inside the outer satellite Hi'iaka
  (from the tides raised by the central body) can be of 
  the order of $10^7\,$W. 
  %about 10 times what is expected from
  %radiogenic heating 
  %\footnote{It would be about several orders of magnitude
%  smaller in case of a synchronous rotation. Then the eccentricity is driving the tides, but a synchronous rotation should not be more probable than another with such non coplanear, eccentric, and irregular orbits.}
  Conversely, taking into
  account the mass, sizes and different Love number, the energy
  dissipated from tides raised by the satellites on Haumea is 
  about~$5 \times 10^9$\,W (somewhat lower than the expected
  radiogenic energy available). \\
  \indent In comparison, the energy needed to crystallize 95\% of
  amorphous ice (starting from an
  equilibrium temperature of $\approx$50\,K) corresponds to an
  increase in temperature of about 40--50\,K
  (with corresponding characteristic times of about $10^8$ and $10^5$ 
  years respectively) 
  \citep{schmitt88}.
  Taking the water ice heat capacity 
  ($C = 2\times10^3$\,J.K$^{-1}$.kg$^{-1}$)
  and the energy found previously,
  one finds a temperature increase rate $\delta T= 3\times 10^{-16}\,$K.s$^{-1}$ for a $10^{19}\,$kg
  satellite and 
  $\delta T= 6\times 10^{-16}\,$K.s$^{-1}$
  for the $\sim 10^{20}\,$kg of
  water ice composing Haumea's crust.
  These values lead to an increase of 50\,K in 
  2 and 5 Gyrs
  for Haumea and Hi'iaka respectively.\\
  \indent This scenario is valid for an energy equally distributed  inside the whole
  volume. If, for some reasons, the energy is mostly dissipated in some
  fraction of the mass, or conducted to the surface, then the increase in
  temperature can be even larger. 
  This could be the case if the surface displays cracks in the ice, 
  were the heating could be more concentrated from the friction occuring
  during the tides; possible cryovolcanism with liquid subsurface
  \citep{desch09} can also be favored by such tidal flexions. \\
  Compared to other systems such as Io or Enceladus, the amount of
  energy involved are very low. However, given the uncertainty of our 
  order of magnitude calculations, it would still be possible, under
  particular conditions only, that tides 
  contribute to the generation of crystalline ice on the satellites surface. 
  Knowledge of the spin vector coordinates of Haumea is required to
  proceed with more specific
  computations for modeling the tides effect,
  the dissipation, heat
  transfer, and ice crystallization. 
  Besides, if the tides are efficient
  to produce crystallisation on the outer satellite, one should
  also expect to have crystalline ice on the inner
  satellite Namaka.\\

\section{Conclusion}

  \indent We presented spectro-imaging observations of (136\,108) Haumea
  obtained in the
  near-infrared [1.6-2.4 $\mu$m] with the integral-field spectrograph
  SINFONI at the ESO VLT.
  The presence of crystalline water ice is confirmed on the surfaces of Haumea
  and Hi'iaka, the largest of the two satellites. Analysis of the spectral bands of water ice and Hapke modeling of our data shows 
  that the surface of Hi'iaka is mainly coated with ``fresh'' ice with larger particle size (20 $\mu$m), supporting a 
  surface less altered than that of Haumea.\\
  Energy sources responsible for the crystallization of the
  water ice are discussed, concluding that radiogenic heating, as well
  as - under very specific conditions - tidal heating, could explain
  this observational result. \\ 
  Improved spectrophotometry of the individual components of the system, and better constrains on Haumea size, 
  shape and spin state, as well as more detailed modelling of the tidal heating,
  are now required to proceed further in this investigation.

\begin{acknowledgements}
      The authors wish to thank the referee Josh Emery, for the pertinent comments he provided to improve the manuscript, H. Hussman (DLR, Berlin), N. Rambaux
      and V. Lainey (IMCCE) for fruitful discussions about the general
      treatment of the tides, as well as the whole ESO-Garching commissioning team for the SINFONI LGS system, in particular Stefan Stroebele, Ronald Donaldon, Sylvain Oberti and Enrico Fedrigo.
\end{acknowledgements}

\end{document}